\documentstyle[12pt]{article}

\newcommand{\beq}{\begin{eqnarray}}
\newcommand{\ene}{\end{eqnarray}}

\newcommand{\F}{\noindent}

\newcommand{\MP}{\medskip}
\newcommand{\BP}{\bigskip}

\begin{document}

\rightline{KIMS-2000-07-10}
\rightline{quant-ph/0007028}
\BP

\vskip12pt

\vskip8pt

\Large

\begin{center}
{\bf Three dimensional time and energy operators and an uncertainty relation}
\vskip10pt

\normalsize
Hitoshi Kitada

Department of Mathematical Sciences

University of Tokyo

Komaba, Meguro, Tokyo 153-8914, Japan

e-mail: kitada@kims.ms.u-tokyo.ac.jp

http://kims.ms.u-tokyo.ac.jp/

\vskip8pt

July 10, 2000

\end{center}
\MP

\vskip10pt

\leftskip=24pt
\rightskip=24pt

\small

\F
{\bf Abstract.} Three dimensional time and energy operators are introduced and an uncertainty relation between them is proved.

\MP

\leftskip=0pt
\rightskip=0pt

\normalsize

\vskip 8pt

Recently, there appeared several papers that introduce three dimensional time in the context of quantum mechanics (see, e.g., \cite{Ch}, \cite{Ti}). The purposes of these papers are different but have similarity in attempting to get better understanding of nature of time and the related areas in physics. We here introduce a new definition of three dimensional time as well as three dimensional energy with the purpose to shed light on the uncertainty relation that holds between them. We will see they are natural extensions of one dimensional time and energy and the uncertainty principle between them holds in quite the same formalism as for the position and momentum operators in usual quantum mechanics unlike the usually conceived derivation of the relation between time and energy. 
\BP

We assume we are given a 3-dimensional coordinate $x=(x_1,x_2,x_3)$ of 3-dimensional Euclidean space $R^3$ and a corresponding 3-dimensional momentum operator $p=(p_1,p_2,p_3)$ conjugate to $x$, i.e. we define
\beq
p_j=\frac{\hbar}{i}\frac{\partial}{\partial x_j},\quad j=1,2,3,
\ene
where $\hbar=h/2\pi$, $h$ being Planck constant. We note in the momentum representation $x_j$ works as a differential operator:
\beq
{\cal F}x_j{\cal F}^{-1}=i\hbar\frac{\partial}{\partial p_j}.
\ene
Here ${\cal F}$ is the Fourier transformation defined by
$$
{\cal F}g(p)=(2\pi\hbar)^{-3/2}\int_{R^3} \exp(-ip\cdot x/\hbar)g(x)dx,
$$
where $p\cdot x=\sum_{j=1}^3 p_j x_j$ is the inner product of the vectors $x$ and $p$.
We assume we are given a time parameter $t$ that takes real values. We then define 3-dimensional time and energy operators $T=(t_1,t_2,t_3)$ and $E=(e_1,e_2,e_3)$ for $t\ne 0$ by\beq
&&t_j=tp_j|p|^{-1},\\
&&e_j=\frac{1}{4t}(|p|x_j+x_j|p|),
\ene
where
$$
|p|=\left(\sum_{j=1}^3 p_j^2\right)^{1/2}
$$
is the positive square root of a nonnegative operator $\sum_{j=1}^3 p_j^2=-\hbar^2\Delta_x$ with $\Delta_x$ being Laplacian with respect to $x$.
We note that these operators $t_j$ and $e_j$ initially defined on the space ${\cal D}={\cal F}^{-1}C_0^\infty(R^3-\{0\})$ can be extended to selfadjoint operators in $L^2(R^3)$, where $C_0^\infty(R^3-\{0\})$ is the space of $C^\infty$-functions with their supports compact in $R^3-\{0\}$ and $L^2(R^3)$ is the space of square integrable functions on $R^3$. Clearly they have dimensions of time and energy, respectively, and for $\pm t>0$, $\pm T=\pm(t_1,t_2,t_3)$ has the same direction as momentum $p$ and satisfies
\beq
\pm|T|=\pm\left(\sum_{j=1}^3 t_j^2\right)^{1/2}=t.
\ene
In this sense our time operator is a modified version of that by Chen \cite{Ch}. Further for $\pm t>0$, $\pm E=\pm(e_1,e_2,e_3)$ has almost the same direction as position $x$, and when $x$ and $p$ denote position and momentum of a scattering particle with mass $m$ whose evolution is governed by a Hamiltonian $H$, we have
\beq
|E|\exp(-itH)g=\left(\sum_{j=1}^3 e_j^2\right)^{1/2}\exp(-itH)g\sim \frac{|p|^2}{2m}\exp(-itH)g
\ene
in $L^2(R^3)$ asymptotically as $t\to\pm \infty$ for a scattering state $g\in L^2(R^3)$, since we have $x_j/t\sim p_j/m$ on $\exp(-itH)g$ as $t\to\pm\infty$ along some sequence $t_k\to\pm\infty$ (see \cite{K1} or \cite{K2}, Theorem 1). Thus these operators can be regarded as a 3-dimensional version of quantum mechanical time and energy. (For a definition of quantum mechanical time of a system with a finite number of particles, see \cite{K1}, \cite{K2}.)

Now our theorem  is the following uncertainty relation between $T$ and $E$.

\F
{\bf Theorem.} Let $f\in {\cal D}\subset L^2(R^3)$ with its $L^2$-norm $\Vert f\Vert=\langle f,f\rangle^{1/2}=1$, where
$$
\langle f,g\rangle =\int_{R^3}f(x)\overline{g}(x)dx
$$
is the inner product of $L^2(R^3)$ with $\overline{g}(x)$ being the complex conjugate of $g(x)$. Let
\beq
&&{\tilde t}_j = \langle t_j f,f\rangle,\quad {\tilde e}_j=\langle e_jf,f\rangle,\nonumber\\
&&{\tilde T}=({\tilde t}_1,{\tilde t}_2,{\tilde t}_3),\quad {\tilde E}=
({\tilde e}_1,{\tilde e}_2,{\tilde e}_3)\nonumber
\ene
be the expectation values of these operators on the state $f$. 
Let the variances of $T$ and $E$ be defined by
\beq
&&\Delta T=\Vert (T-{\tilde T})f\Vert=\left(\sum_{j=1}^3\Vert(t_j-{\tilde t}_j)f\Vert^2\right)^{1/2},\nonumber\\
&&\Delta E=\Vert(E-{\tilde E})f\Vert=\left(\sum_{j=1}^3\Vert(e_j-{\tilde e}_j)f\Vert^2\right)^{1/2}.\nonumber
\ene
Then we have the uncertainty relation:
\beq
\Delta T\Delta E\ge \frac{\hbar}{2}.
\ene
\BP

\F
{\it Proof.} We note using Schwarz inequality
\beq
\Delta T\Delta E&=&\Vert (T-{\tilde T})f\Vert\Vert(E-{\tilde E})f\Vert\nonumber\\
&\ge& \left|\sum_{j=1}^3\langle (t_j-{\tilde t}_j)f,(e_j-{\tilde e}_j)f\rangle \right|\nonumber\\
&=&\left|\sum_{j=1}^3\{\langle t_jf,e_jf\rangle -{\tilde t}_j{\tilde e}_j\}\right|.\nonumber
\ene
Let $\mbox{Im}\ z$ denote the imaginary part of a complex number $z$. Then noting ${\tilde t}_j$ and ${\tilde e}_j$ are real numbers, we have
\beq
\Delta T\Delta E\ge
 \left|\mbox{Im}\sum_{j=1}^3\{\langle t_jf,e_jf\rangle -{\tilde t}_j{\tilde e}_j\}\right|
=\left| \mbox{Im}\sum_{j=1}^3\langle t_jf,e_jf\rangle \right|
=\left| \sum_{j=1}^3\frac{1}{2}\langle [t_j,e_j]f,f\rangle \right|.
\ene
Here we compute
\beq
4[t_j,e_j]f&=&[p_j|p|^{-1},(|p|x_j+x_j|p|)]f\nonumber\\
&=&([p_j,x_j]+|p|^{-1}p_jx_j|p|-|p|x_jp_j|p|^{-1})f\nonumber\\
&=&(i^{-1}\hbar+|p|^{-1}(i^{-1}\hbar+x_jp_j)|p|-|p|x_jp_j|p|^{-1})f\nonumber\\
&=&(2i^{-1}\hbar+|p|^{-1}x_jp_j|p|-|p|x_jp_j|p|^{-1})f.\nonumber
\ene
Using equality (2) above we have
\beq
[x_j,|p|]=i\hbar p_j|p|^{-1}.
\ene
Thus
\beq
4[t_j,e_j]&=&2i^{-1}\hbar+|p|^{-1}([x_j,|p|]+|p|x_j)p_j+([x_j,|p|]-x_j|p|)|p|^{-1}p_j\nonumber\\
&=&2i^{-1}\hbar+2i\hbar p_j^2|p|^{-2}.\nonumber
\ene
Then
\beq
\sum_{j=1}^3[t_j,e_j]=i^{-1}\hbar.\nonumber
\ene
From this, (8) and $\Vert f\Vert=1$ we have
\beq
\Delta T\Delta E\ge \left|\frac{1}{2}\frac{\hbar}{i}\right|=\frac{\hbar}{2}.\nonumber
\ene
$\Box$
\BP

\end{document}